\documentclass[%
 reprint,
 superscriptaddress,
 amsmath,amssymb,
 aps,
]{revtex4-2}

\usepackage{graphicx}
\usepackage{dcolumn}
\usepackage{bm}

\usepackage{siunitx}
\usepackage{chemformula}
\usepackage{comment}
\usepackage{pifont}
\newcommand{\cmark}{\ding{51}}
\newcommand{\xmark}{\ding{55}}
\usepackage{soul} 

\begin{document}

\title{Piezoelectrically actuated high-speed spatial light modulator for visible to near-infrared wavelengths}

\author{Tom Vanackere}
\altaffiliation{These authors contributed equally to this work; \\ tomvac96@mit.edu, arturh@mit.edu}
\affiliation{Massachusetts Institute of Technology, Cambridge, MA, USA}
\affiliation{Ghent University - imec, Ghent, Belgium}
\author{Artur Hermans}
\altaffiliation{These authors contributed equally to this work; \\ tomvac96@mit.edu, arturh@mit.edu}
\affiliation{Massachusetts Institute of Technology, Cambridge, MA, USA}
\affiliation{Ghent University - imec, Ghent, Belgium}
\author{Ian Christen}
\affiliation{Massachusetts Institute of Technology, Cambridge, MA, USA}
\author{Christopher Panuski}
\affiliation{Massachusetts Institute of Technology, Cambridge, MA, USA}
\author{Mark Dong}
\affiliation{Massachusetts Institute of Technology, Cambridge, MA, USA}
\affiliation{The MITRE Corporation, Bedford, MA, USA}
\author{Matthew Zimmermann}
\affiliation{The MITRE Corporation, Bedford, MA, USA}
\author{Hamza Raniwala}
\affiliation{Massachusetts Institute of Technology, Cambridge, MA, USA}
\author{Andrew J. Leenheer}
\affiliation{Sandia National Laboratories, Albuquerque, NM, USA}
\author{Matt Eichenfield}
\affiliation{Sandia National Laboratories, Albuquerque, NM, USA}
\affiliation{University of Arizona, Tucson, AZ, USA}
\author{Gerald Gilbert}
\affiliation{The MITRE Corporation, Princeton, NJ, USA}
\author{Dirk Englund}
\email{englund@mit.edu}
\affiliation{Massachusetts Institute of Technology, Cambridge, MA, USA}

\date{\today}

\begin{abstract}
Advancements in light modulator technology have been driving discoveries and progress across various fields. The problem of large-scale coherent optical control of atomic quantum systems---including cold atoms, ions, and solid-state color centers---presents among the most stringent requirements. This motivates a new generation of high-speed large-scale modulator technology with the following requirements: (R1) operation at a design wavelength of choice in the visible (VIS) to near-infrared (NIR) spectrum, (R2) a scalable technology with a high channel density ($>\SI{100}{mm^{-2}}$), (R3) a high modulation speed ($>\SI{100}{\mega\hertz}$), and (R4) a high extinction ratio ($>\SI{20}{\decibel}$). To fulfill these requirements, we introduce a modulator technology based on piezoelectrically actuated silicon nitride resonant waveguide gratings fabricated on \SI{200}{\milli\metre} diameter silicon wafers with CMOS-compatible processes. We present a proof-of-concept device with $4\times4$ individually addressable $\SI{50}{\micro\metre}\times\SI{50}{\micro\metre}$ pixels or channels, each containing a resonant waveguide grating with a $\sim\SI{780}{\nano\metre}$ design wavelength, supporting $>\SI{100}{\mega\hertz}$ modulation speeds, and a spectral response with $>\SI{20}{\decibel}$ extinction. 
\end{abstract}

\maketitle

Devices that have the ability to control multiple optical modes at high speeds are revolutionizing many fields of science and technology, including laser ranging \cite{Park:21}, quantum information processing \cite{Ebadi:21}, imaging \cite{Nikolenko:08,Yoon:20}, optogenetics \cite{Ronzitti:17}, artificial intelligence \cite{Zuo:21,Bernstein:22}, and augmented and virtual reality (AR/VR) \cite{Chang:20,Zhan:20}. In particular, the availability of multimodal and/or high-speed light modulators has been instrumental in the rapid advancement of atom-based quantum technology over the last years \cite{Wright:19,Ebadi:21,Christen:22,Graham:22,Menssen:23}. This motivates the development of a new modulator technology, meeting requirements (R1)-(R4), to continue driving progress. Many optical transitions in atoms or atom-like emitters in solids lie in the visible (VIS) to near-infrared (NIR) wavelength range, hence requirement (R1). VIS to NIR wavelength operation is also essential for bioimaging \cite{Lichtman:05,Yoon:22}, optogenetics \cite{Ronzitti:17}, and AR/VR \cite{Chang:20}. Requirement (R2), a scalable technology with a high channel density, stems from the need to individually control hundreds to thousands of atoms to unlock quantum technology's full potential. High modulation speeds (R3) and high extinction ratios (R4) are required to optically control the state of atoms or atom-like emitters with high fidelity in a short time compared to the lifetime of the state \cite{Menssen:23}. 
A high-extinction and high-channel-density optical control technology acting as an ultrafast spatial light modulator is also an immediate benefit to the varied fields of research which already rely on devices with thousands to millions of times slower modulation rates.

\begin{table*}[hbt!]
\caption{\label{tab:comparison_modulator_tech}Comparison of multichannel light modulator technologies in terms of requirements (R1)-(R4). }
\begin{ruledtabular}
\begin{tabular}{ccccc}
 Technology
 &R1&R2&R3&R4\\ \hline 
 Liquid-Crystal-on-Silicon (LCoS) \cite{McKnight:94,Zhang:14}
 &\cmark&\cmark &\xmark & \cmark \\
 Digital Micromirror Device (DMD)\footnote{MEMS device} \cite{Hornbeck:97}
 &\cmark&\cmark &\xmark & \cmark \\ 
 Grating Light Valve (GLV)$^\mathrm{a}$ \cite{Bloom:97}
 &\cmark&\cmark &\xmark & \cmark \\
 Atom-control Photonic Integrated Circuit (APIC)\footnote{SiN PIC with AlN piezoelectric actuators} \cite{Menssen:23}
 &\cmark&\xmark &\cmark & \cmark \\
 Electro-optic silicon-organic metasurface \cite{Benea-Chelmus:22}
 &\xmark&\xmark &\cmark & \xmark \\
 Silicon photonic crystal cavities with free-carrier modulation \cite{Panuski:22}
 &\xmark&\cmark &\cmark & - \\
 Technology presented in this work 
 &\cmark&\cmark &\cmark & \cmark \\
\end{tabular}
\end{ruledtabular}
\end{table*}

\begin{figure*}[!ht]
\centering
\includegraphics{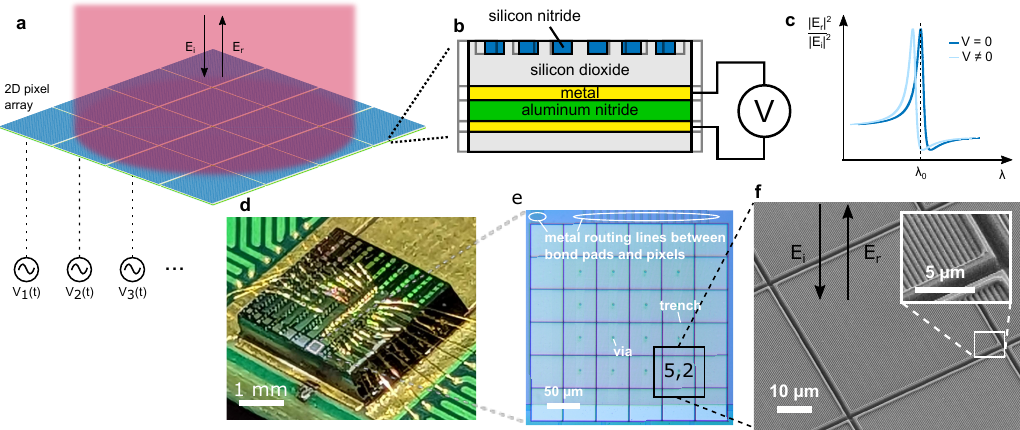}
\caption{\label{fig:concept_and_fabricated_device} Concept and device images. \textbf{a,b,} Concept figure of our spatial light modulator (SLM) technology based on piezoelectrically actuated SiN resonant waveguide gratings. The piezoelectric actuators are based on AlN thin films. \textbf{c,} Device operation principle. Applying a voltage shifts the reflection spectrum. \textbf{d,} Chip with $4\times4$ SLM wire bonded to a printed circuit board. See Methods for packaging details. \textbf{e,} Microscope image of $4\times4$ SLM with a \SI{50}{\micro\metre} pixel pitch and \SI{1}{\micro\metre} wide trenches in between pixels. \textbf{f,} Scanning electron microscope (SEM) image showing a grating surrounded by trenches.}
\end{figure*}

\begin{figure*}[!ht]
\centering
\includegraphics{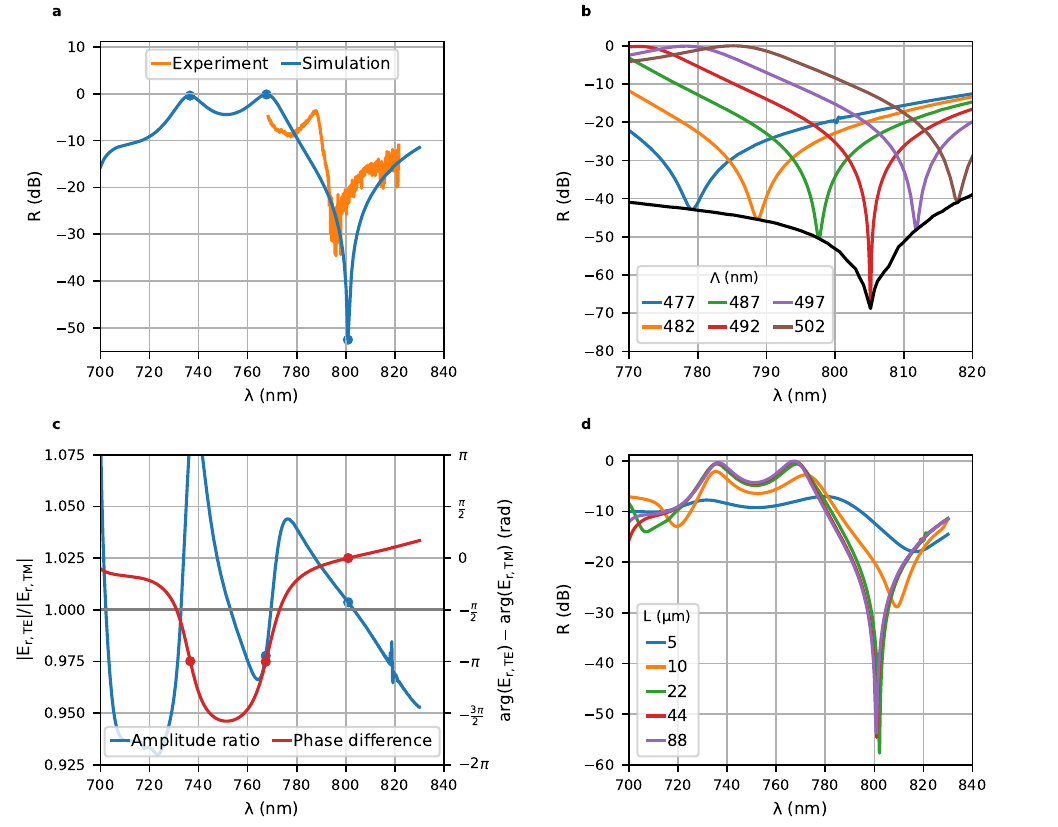}
\caption{\label{fig:Design_device} Device design, optical. \textbf{a,} Simulated reflection spectrum compared to the measured spectrum, with $\mathrm{R (dB)} = 10 \times\log_{10} \Big( \frac{|\mathrm{E_{out}(\lambda)}|^2}{|\mathrm{E_{ref}(\lambda)}|^2}\Big)$. \textbf{b,} Influence of the grating pitch $\mathrm{\Lambda}$ on the reflection spectrum. These simulations are all for a duty cycle of \SI{63}{\percent}. \textbf{c,} The simulated amplitude ratio and phase difference between TE and TM polarized reflected waves. The dots indicate the wavelengths at which there is a $0$ and $-\pi$ phase differences, corresponding to the minima and maxima in \textbf{a}.
\textbf{d,} Simulation of reflection spectrum for different grating sizes $\mathrm{L}$. The $1/e$ field full width is kept at \SI{45}{\percent} of the grating size.}
\end{figure*}

\begin{figure*}[!ht]
\centering
\includegraphics{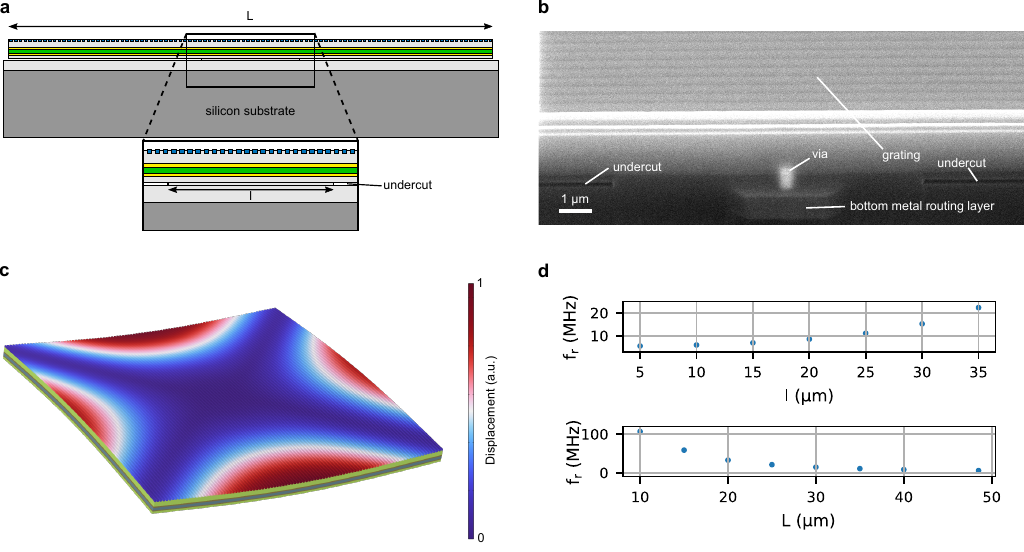}
\caption{\label{fig:Design_device_2} Device design, mechanical. \textbf{a,} Schematic cross-section of a released device with grating size $\mathrm{L}$. Released devices are undercut and anchored to the substrate with a silicon dioxide pillar, with diameter $\mathrm{l}$, in the middle of the grating. \textbf{b,} SEM image of a cross-section of a released device (at an angle of \ang{57.5}). We used focused ion beam milling (FIB) to create a hole in the device so we could image the cross-section. This is an image of the middle of a grating, where it is anchored to the substrate. \textbf{c,} Simulated mechanical mode with an eigenfrequency $\mathrm{f_r}$ of \SI{5.9}{\mega\hertz} for a released device with $\mathrm{L}=\SI{49}{\micro\metre}$ and $\mathrm{l}=\SI{10}{\micro\metre}$. \textbf{d,} Simulated eigenfrequency $\mathrm{f_r}$ as a function of pillar diameter $\mathrm{l}$ for grating size $\mathrm{L}=\SI{49}{\micro\metre}$ (top) and simulated $\mathrm{f_r}$ as a function of $\mathrm{L}$ for $\mathrm{l}=\SI{5}{\micro\metre}$ (bottom).}
\end{figure*}

Table \ref{tab:comparison_modulator_tech} compares state-of-the-art multichannel light modulator technologies in terms of requirements (R1)-(R4). None of the existing technologies meets all requirements. Spatial light modulators (SLMs) operating in the VIS to NIR range based on microelectromechanical systems (MEMS) \cite{Hornbeck:97,Bloom:97,Tzang:19} and liquid crystals \cite{McKnight:94,Zhang:14} are commercially available, but their modulation rates are limited. In the IR wavelength range, alternative SLM technologies with $>\SI{1}{\mega\hertz}$ modulation rates have been explored based on electrical gating of graphene \cite{Fan:16,Zeng:18}, the Pockels effect in organic electro-optic molecules \cite{Benea-Chelmus:21,Sun:21,Benea-Chelmus:22} and dielectric thin films \cite{Smolyaninov:19}, the quantum-confined Stark effect in III-V semiconductor quantum-well-based material \cite{Worchesky:96,Lee:14}, and charge carrier effects in indium tin oxide (ITO) \cite{Huang:16} and silicon \cite{Shuai:17,Panuski:22}. The refractive index changes associated with these fast effects tend to be small, thus long interaction lengths are needed to realize devices with a high modulation efficiency. To create compact, free-space-coupled devices, one therefore often relies on optically resonant structures, such as metasurfaces \cite{Fan:16,Zeng:18,Sun:21,Lee:14,Huang:16}, Fabry-Perot cavities \cite{Worchesky:96}, resonant waveguide gratings \cite{Benea-Chelmus:21,Benea-Chelmus:22,Shuai:17}, or photonic crystal cavities \cite{Panuski:22}. 

To fulfill requirements (R1)-(R4), we introduce an SLM technology based on piezoelectrically actuated silicon nitride (SiN) resonant waveguide gratings (Fig. \ref{fig:concept_and_fabricated_device}). The piezoelectric actuators induce strain in the gratings, resulting in a shift of the resonant wavelengths and thereby realizing  free-space-coupled modulators. This technology not only enables SLMs with $\mathcal{O}(\si{\mega\hertz})$ to $\mathcal{O}(\si{\giga\hertz})$ modulation rates, it also operates in the VIS to NIR wavelength range. Moreover, our devices are fabricated on \SI{200}{\milli\metre} diameter silicon wafers using deep ultraviolet (UV) optical lithography and CMOS-compatible (complementary-metal-oxide-semiconductor-compatible), low-temperature processes (see Methods for fabrication details). Hence, we can fabricate the piezoelectric actuators and resonant waveguide gratings directly on top of electronics to drive the devices \cite{Lee:13} or on (multiple) metal layers to enable large-scale electrical interconnects. These integration schemes are advantageous for the commercial production of large-scale SLMs. 

Our work builds on a platform developed at Sandia National Laboratories consisting of a SiN optical waveguide layer on top of an aluminum-nitride-based (AlN) piezoelectric actuation layer \cite{Stanfield:19}. While previous works on this platform focused on photonic integrated circuits \cite{Stanfield:19,Dong:22,Dong:22b,Menssen:23}, here we demonstrate densely packed free-space-coupled modulators arranged in a $4\times4$ grid with a high fill factor of $\sim\SI{96}{\percent}$ as a proof of concept for scalable SLM technology (Fig. \ref{fig:concept_and_fabricated_device}). To boost the modulation efficiency of our devices we not only make use of optically resonant devices, namely resonant waveguide gratings \cite{Quaranta:18}, but we also take advantage of the mechanical resonances of our pixels, which greatly enhances the piezoelectrically induced displacement \cite{Li:21}.

\begin{figure*}[!ht]
\centering
\includegraphics{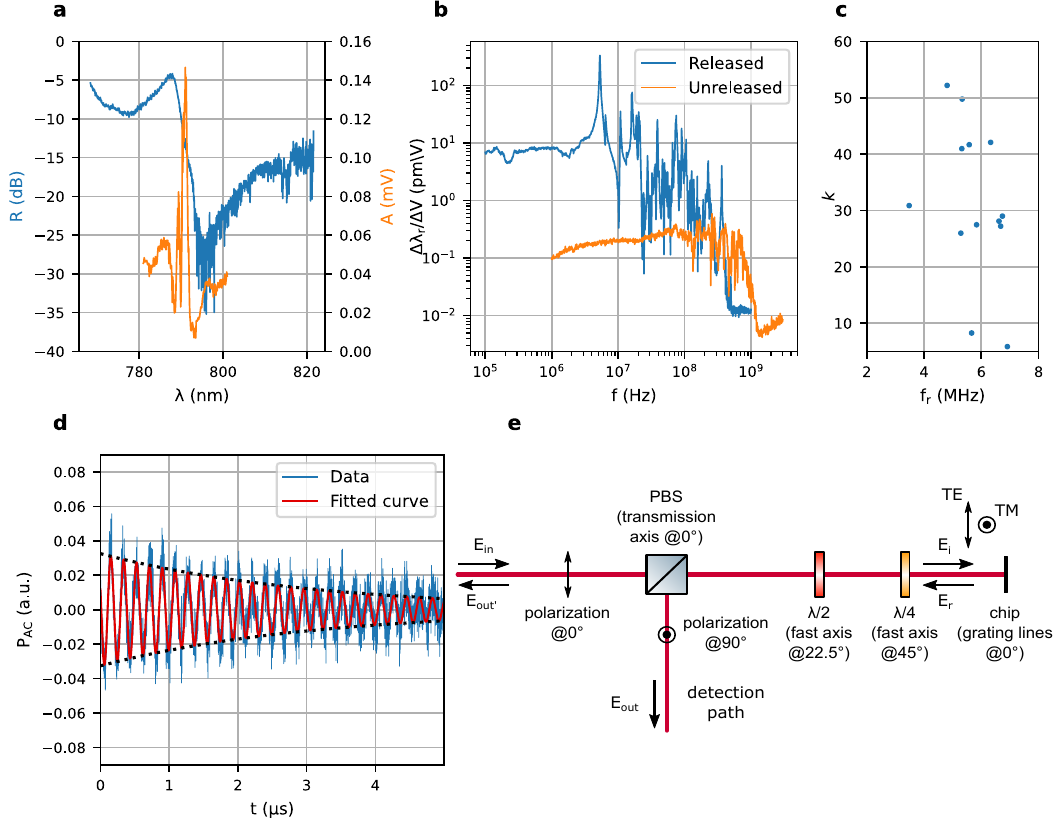}
\caption{\label{fig:measurement_results}Device characterization. \textbf{a,} The reflection spectrum of the grating (blue) overlapped with the modulation amplitude $\mathrm{A}$ as a function of wavelength (orange), measured on an unreleased device at a fixed AC frequency $\mathrm{f}$ of \SI{20}{\mega\hertz} for a peak-to-peak voltage of \SI{10}{\V}. \textbf{b,} Wavelength shift $\mathrm{\Delta\lambda_r/\Delta V}$ versus AC frequency $\mathrm{f}$ for a released and unreleased device. We used a detector with a \SI{3}{\decibel} bandwidth of DC-\SI{400}{\mega\hertz} for the released device and \num{5}-\SI{1000}{\mega\hertz} for the unreleased device. The inset shows the simulated mechanical mode at the resonance frequency $\mathrm{f_r}$ of $\sim\SI{5}{\mega\hertz}$. \textbf{c,} Measurement of the mechanical resonance frequencies $\mathrm{f_r}$ of the strong resonance around \SI{5}{\mega\hertz} and the associated enhancement factor $k$ for each pixel. 13 of the 16 pixels were functioning in this device. \textbf{d,} Ringdown measurement of the mechanical mode at $\sim\SI{5}{\mega\hertz}$ of a released device. For this measurement we first apply an AC frequency at resonance, then we switch it off and observe how the AC-coupled photodiode signal $\mathrm{P_{AC}}$ decays over time $\mathrm{t}$ on an oscilloscope. Fitting gives a ringdown time $\mathrm{\tau}$ of \SI{3.1}{\micro\second}. \textbf{e,} Schematic depiction of our cross-polarized light setup.}
\end{figure*}

\section{\label{sec:results}Results}
\subsection*{\label{sec:design}Design}
We designed our SiN resonant waveguide gratings to operate at a wavelength $\mathrm{\lambda}$  of $\sim\SI{780}{\nano\metre}$ (R1), with a pitch $\mathrm{\Lambda}$ of \SI{0.490}{\micro\metre} for a SiN thickness of \SI{0.3}{\micro\metre}. The simulated reflection spectrum of our gratings, $\mathrm{R (dB)} = 10 \times\log_{10} \Big( \frac{|\mathrm{E_{out}(\lambda)}|^2}{|\mathrm{E_{ref}(\lambda)}|^2}\Big)$, in Fig. \ref{fig:Design_device}a shows a steep dip at a wavelength of $\sim\SI{800}{\nano\metre}$ with an extinction ratio $>\SI{50}{\decibel}$ (R4) with respect to the reflection at $\sim\SI{770}{\nano\metre}$. We test our devices in a cross-polarized light setup, as schematically depicted in Fig. \ref{fig:measurement_results}e (see supplementary information for more details). Our simulation of the reflection spectrum captures the cross-polarized testing method (see Methods for more information). The steep dip in the reflection spectrum results from the phase difference between TE and TM polarized reflected light shifting from 0 to $\pi$ while the reflection amplitude of TE and TM fields remain similar (Fig. \ref{fig:Design_device}c), where TE polarized light corresponds to the electric field being parallel to the grating lines (Fig. \ref{fig:measurement_results}e). When the TE and TM polarized light waves reflected off the grating are in phase with equal amplitude, the reflected light is polarized at \ang{0} before the polarizing beamsplitter and no light is reflected towards the detector. For a phase difference of $\pi$ all light is reflected. Notably, the effective interferometer governing this effect is contained within $\sim\SI{1}{\micro\metre}$ of thickness of the device, requiring no long-pathlength stabilization.

A decrease in grating pitch blue-shifts the wavelength of the dip in the reflection spectrum (Fig. \ref{fig:Design_device}b). A pitch of \SI{492}{\nano\metre} gives a maximum extinction ratio of $>\SI{60}{\decibel}$. The reduced extinction ratio at different pitches results from the ratio between the TE and TM reflection amplitudes being further away from 1 where the phase difference is 0. 

Our simulations take into account the finite size of the gratings and the incident light beam. While our fabricated and tested devices have a grating size $\mathrm{L}$ of \SI{49}{\micro\metre} (Fig. \ref{fig:concept_and_fabricated_device}e), simulations indicate that we could reduce the size down to $\sim\SI{20}{\micro\metre}$ without significant changes in extinction ratio and the overall shape of the reflection spectrum (Fig. \ref{fig:Design_device}d), resulting in even higher channel densities (R2).

Fig. \ref{fig:Design_device_2} shows the mechanical properties of the device. After performing the undercut the pixel of length $L$ is supported by a pillar of size $l$ underneath, which also contains the vias to the electrodes. A SEM cut shows the cross section of a pixel, Fig. \ref{fig:Design_device_2}b. Simulations in COMSOL of the mechanical properties of such a pixel show a symmetrical mode with a eigenfrequency frequency close to the one measured in experiments Fig. \ref{fig:measurement_results}b. Fig. \ref{fig:Design_device_2}d shows how the eigenfrequency of the mode changes based on the pixel size $L$ and the pillar size $l$. Using this one could design the gratings for enhanced functionality at a desired frequency.

\subsection*{\label{sec:characterization}Characterization}
Fig. \ref{fig:measurement_results}a and \ref{fig:Design_device}a show the measured reflection spectrum of a grating and its agreement with simulations, with a measured extinction ratio $>\SI{20}{\decibel}$ (R4). We measure the modulation amplitude $\mathrm{A}$ and speed of the packaged devices by reflecting laser light off a pixel and detecting the reflected light (with electric field $\mathrm{E_{out}}$) on a fast photodiode (Fig. \ref{fig:measurement_results}e), which is connected to an electrical spectrum analyzer (ESA). When applying an AC voltage to the illuminated pixel, a peak appears on the ESA at the applied AC frequency $\mathrm{f}$, as anticipated for our piezoelectrically actuated device. The height of this peak, i.e. the modulation amplitude $\mathrm{A}$, depends on the laser wavelength. Fig. \ref{fig:measurement_results}a overlaps the reflection spectrum with the modulation amplitude as a function of wavelength $\mathrm{\lambda}$. The peak of the modulation amplitude occurs where the slope in the reflection spectrum is steepest ($\sim\SI{791}{\nano\metre}$). 

Fig. \ref{fig:measurement_results}b compares the modulation amplitude and speed for two different types of devices, released and unreleased. Released devices are undercut and anchored to the substrate with a \SI{10}{\micro\metre} wide silicon dioxide pillar in the middle of the pixel (see Methods for fabrication details and supplementary information for an SEM image of a device cross-section). Unreleased devices have no undercut. The goal of releasing is to make the devices more mechanically compliant, resulting in lower drive voltages for the same level of induced strain. Additionally, released devices have high-quality-factor mechanical resonances. Resonant enhancement reduces the required drive voltage even further. Fig. \ref{fig:measurement_results}b shows the wavelength shift $\mathrm{\Delta\lambda_r/\Delta V}$ of the reflection spectrum versus the applied AC frequency $\mathrm{f}$ for a released and unreleased grating. We calculate the wavelength shift from the measured modulation amplitude and the slope of the reflection spectrum. The numerous peaks and troughs in Fig. \ref{fig:measurement_results}b are attributed to mechanical resonances \cite{Dong:23, Wen:23}. The more compliant nature of the released devices results in low-frequency wavelength shifts that are more than an order of magnitude larger compared to the unreleased devices. It also leads to the appearance of mechanical resonances at lower frequencies. Finite element analysis confirms the existence of mechanical resonances and identifies the mechanical mode at $\mathrm{f_r} \approx \SI{5}{\mega\hertz}$ for the released device (inset in Fig. \ref{fig:measurement_results}b). The resonance at $\mathrm{f_r} \approx \SI{5}{\mega\hertz}$ gives a large wavelength shift of \SI{4e2}{\pm\per\V}, which corresponds to an enhancement of a factor $k = \mathrm{\frac{(\Delta\lambda_r/\Delta V)_{f_r}}{(\Delta\lambda_r/\Delta V)_{DC}}} \approx 50$ compared to the low-frequency shift. Up to a frequency of $\sim\SI{100}{\mega\hertz}$, the data shows wavelength shifts larger than the low-frequency response for the released device (R3). The stiffer nature of the unreleased device shifts the resonances to higher frequencies. Up to $\sim\SI{700}{\mega\hertz}$, wavelength shifts are comparable to the low-frequency response. The frequency response measurements are limited by the photodetector bandwidths (DC-\SI{400}{\mega\hertz} for the released device and \num{5}-\SI{1000}{\mega\hertz} for the unreleased device measurement) and the \SI{900}{\mega\hertz} bandwidth of the voltage amplifiers on the printed circuit boards (see Methods). 

Fig. \ref{fig:measurement_results}d shows the ringdown measurement of the mechanical mode at $\sim\SI{5}{\mega\hertz}$ for a released device. Fitting the data gives a ringdown time $\mathrm{\tau}$ of \SI{3.1}{\micro\second}. This corresponds to a mechanical Q-factor of $\mathrm{Q_m} = \mathrm{\frac{\omega_m \tau}{2}} = 52$ ($\mathrm{\omega_m} = 2\pi \times \SI{5.3}{\mega\hertz}$) \cite{Thompson:08}. This is in good agreement with the enhancement factor of $\sim50$ observed in the frequency response measurement, as expected for a harmonic oscillator. 

Pixel-to-pixel variations lead to differences in resonance frequencies and enhancement factors. For the pixels in our $4 \times 4$ device, we determined the mechanical resonance frequencies $\mathrm{f_r}$ of the strong resonance around \SI{5}{\mega\hertz} and the associated enhancement factor $k$ (Fig. \ref{fig:measurement_results}c). The average resonance frequency is \SI{5.7}{\mega\hertz} (standard deviation of \SI{0.9}{\mega\hertz}) and the average enhancement factor is 32 (standard deviation of 13). Established post-fabrication trimming techniques can be used to tune up the resonances \cite{Liu:23,Efimovskaya:20}.

\section{\label{sec:discussion}Discussion}
One of the main challenges ahead is to reduce the voltages required to achieve high-extinction-ratio modulation. We see several ways to achieve this. Firstly, we can use a piezoelectric material with an enhanced response. Scandium aluminum nitride, for instance, has a piezoelectric response up to 5 times larger compared to AlN \cite{Akiyama:09}. Secondly, instead of resonant waveguide gratings, we can use other compact, optically resonant structures with narrower optical linewidths. Vertically coupled photonic crystal cavities with high quality factors and small mode volumes are a promising candidate. These have been demonstrated in silicon with linewidths $<\SI{10}{\pico\metre}$ at a wavelength of \SI{1.6}{\micro\metre} \cite{Panuski:22}. Silicon nitride photonic crystal cavities with linewidths down to \SI{0.1}{\nano\metre} at \SI{760}{\nano\metre} have been reported \cite{Fryett:18}, though these were not vertically coupled. 

We see many potential applications for this SLM technology. Depending on the target application, we can configure our modulator technology in different ways. For instance, driving the modulators at a strong mechanical resonance generates strong sidebands to the incident laser light (the sidebands could be resonant with an atomic transition). We can use different mechanical resonances to generate sidebands at different frequencies or engineer the pixels to alter the mechanical resonances, e.g. by varying the grating size or the depth of the undercut. Laser trimming \cite{Panuski:22,Liu:23} or focused ion beam milling \cite{Efimovskaya:20} can alter the resonance frequencies post-fabrication.

Predictable time-varying and even periodically repeating light patterns can be generated \cite{Dong:23} by driving each of the pixels at a particular mechanical resonance frequency, enabling low-voltage driving, with a particular phase. The generated pattern will depend on the chosen frequencies and phases. This has applications in atom control, display technology, and laser ranging.

Our modulator technology could even generate arbitrary waveforms requiring the full device bandwidth (i.e. not only driving at resonance frequencies) by using signal equalization techniques to compensate for the non-flat frequency response \cite{Wen:23}. 

In conclusion, we have presented a scalable high-channel-density, high-speed modulator technology operating in the visible to near-infrared spectrum. This technology has the potential to drive progress across various fields in science and technology, including large-scale coherent optical control of atomic quantum systems.

\bibliography{biblio}

\clearpage
\section*{\label{sec:methods}Methods}

\subsection*{\label{sec:fab}Fabrication}
Our devices consist of a silicon nitride (SiN) optical waveguide layer on top of an aluminum-nitride-based (AlN) piezoelectric actuation layer with a silicon dioxide optical buffer layer in between \cite{Stanfield:19,Dong:22,Dong:22b}. The devices are fabricated on \SI{200}{\milli\metre} diameter silicon wafers at Sandia National Laboratories using deep ultraviolet (UV) optical lithography. Low-temperature, CMOS-compatible processes are used for thin-film deposition, namely plasma-enhanced chemical vapor deposition (PECVD; for amorphous silicon, SiN and oxide deposition) and physical vapor deposition (PVD; for AlN and metal deposition). There are three metal layers in the stack to enable device actuation. From bottom to top, these are: a routing layer which we use to connect each of the pixels to bond pads, a bottom electrode layer, and a top electrode layer. In between the two electrode layers there is the AlN layer; together these comprise the piezoelectric actuation layer. The bottom electrode layer is separated from the routing layer by a silicon dioxide buffer. The three metal layers are connected by vias. Throughout the process flow, chemical-mechanical polishing (CMP) is used for planarization. In between the bottom electrode layer and routing layer there is also a patterned amorphous silicon layer to enable undercutting of devices. After having processed the entire device stack, a release etch in XeF$_2$ gas removes the amorphous silicon to create undercut devices. The XeF$_2$ gas can reach the amorphous silicon layer through deep, etched trenches. 

\subsection*{\label{sec:packaging}Packaging}
We dice the wafers into chips and wire bond them to printed circuit boards (Fig. \ref{fig:concept_and_fabricated_device}d). The devices with $4\times4$ pixels have 17 bond pads, 1 pad for a shared ground and 16 pads for the signals applied to each of the pixels. Each of our PCBs contains 16 operational amplifiers (OPAMPs; Texas Instruments THS3491) to individually address 16 pixels. The OPAMPs provide a voltage gain of 5, have a \SI{900}{\mega\hertz} bandwidth, and a $\pm\SI{16}{\volt}$ supply voltage. Ganged microscale RF plugs (Samtec) connect the PCBs to signal generators.

\subsection*{\label{sec:sims}Simulation}
We use Ansys Lumerical's 2D FDTD solver for optical simulations. We run the simulation for two orthogonal polarizations. Afterwards, we use transfer matrix multiplication to implement the beamsplitter and wave plates together with the simulations of both polarizations. Our simulation takes into account the Gaussian shape of the excitation and collected beams. To directly compare the simulation with the measurement, we implement the same normalization method. We perform a simulation without the grating structure to simulate the effect of reflection from the aluminum layer with the silicon dioxide and SiN layers on top ($\mathrm{E_{ref}}$). The resulting simulation is able to capture the entire measurement setup to accurately predict the results.

\begin{acknowledgments}
We acknowledge funding from MITRE's Quantum Moonshot program, the Defense Advanced Research Projects Agency (ONISQ program), the NSF Center for Quantum Networks, and the Air Force Research Laboratory. This work is supported by a collaboration between the US DOE and other Agencies. This material is based upon work supported by the U.S. Department of Energy, Office of Science, National Quantum Information Science Research Centers, Quantum Systems Accelerator. T. V. and A. H. thank the Research Foundation - Flanders (FWO) for financial support (PhD fellowship 11F5322N and postdoctoral fellowship 1258423N). D.E. acknowledges partial support from Honda Research Institute USA, Inc., and from the NSF Eager program. C.L.P. was supported by the Hertz Foundation Elizabeth and Stephen Fantone Family Fellowship. Simulations were supported in part by Army Research Office grant W911NF-20-1-0084 supervised by M. Gerhold. Sandia National Laboratories is a multimission laboratory managed and operated by National Technology \& Engineering Solutions of Sandia, LLC, a wholly owned subsidiary of Honeywell International Inc., for the U.S. Department of Energy’s National Nuclear Security Administration under contract DE-NA0003525. This paper describes objective technical results and analysis. Any subjective views or opinions that might be expressed in the paper do not necessarily represent the views of the U.S. Department of Energy or the United States Government.
\end{acknowledgments}

\newpage
\onecolumngrid
\newpage
\appendix
\section{Piezoelectrically actuated high-speed spatial light modulator for visible to near-infrared wavelengths: supplementary information}
\subsection{\label{sec:meas}Measurement setup}
Linearly polarized light from a Ti:sapphire laser is sent through a polarizing beamsplitter and rotated \ang{45} through the use of a half-wave plate. The light then impinges perpendicularly onto the resonant waveguide gratings. The TE and TM polarizations in the light experience a different amplitude and phase change during the reflection such that when the light passes through the wave plates again, the resulting polarization has changed. When the reflected light then hits the polarizing beamsplitter, part of it is reflected to the detector rather than transmitted back to the laser. The detector is a high-speed avalanche photodiode (APD). By adjusting the voltage over the piezoelectric layer, we can change the amplitude and phase change both polarizations experience during reflection and control the portion of the light that is reflected by the beamsplitter. 

For the measurement of the reflection spectrum, we replaced the Ti:sapphire laser with a superluminescent diode (SLD), and sent the light to a spectrometer instead of an APD. To normalize the measurement of the reflection spectrum, we also measured the reflection spectrum from a uniform area of the chip that does not include the grating structure, where the light is simply reflected back by the aluminum electrode layer.

\begin{figure}[h]
\centering
\includegraphics{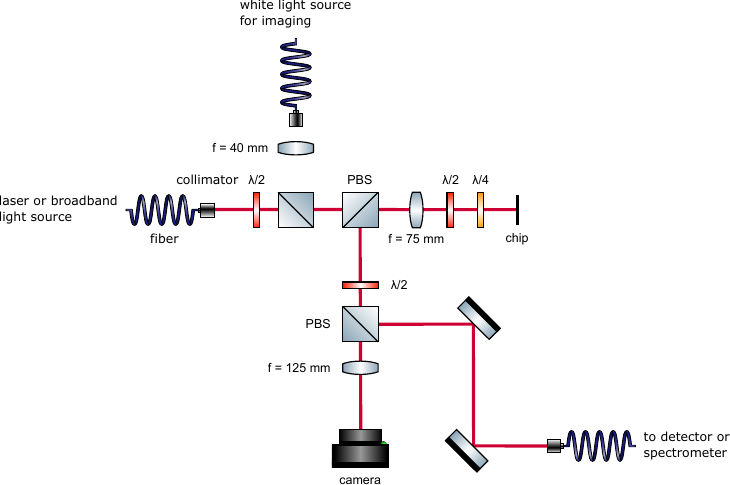}
\caption{Schematic of the setup we used for characterization of our spatial light modulators. PBS: polarizing beamsplitter, $\lambda/2$: half-wave plate, $\lambda/4$: quarter-wave plate.}
\label{fig:setup}
\end{figure}
\newpage

\subsection{Field profile}

\begin{figure}[h]
\centering
\includegraphics{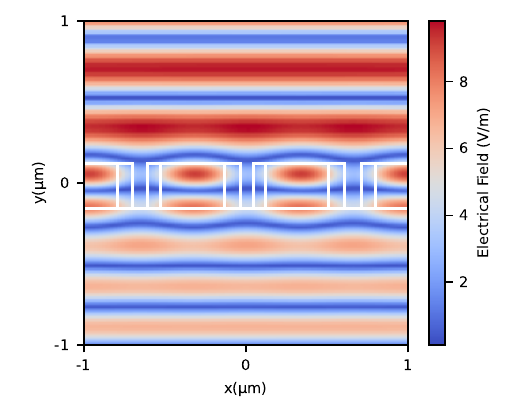}
\caption{Field profile of the TE polarized light in the structure at a wavelength of \SI{800}{\nano\metre}. The white lines depict the edges of the silicon nitride and silicon dioxide structures. The grating is only partly filled with oxide due to deposition limitations.}
\end{figure}

\end{document}